\documentclass[12pt]{article}
\usepackage{graphicx}
\usepackage[margin=1in]{geometry}
\usepackage{xcolor}
\usepackage{setspace}
\doublespace
\usepackage{url}
\usepackage{xr}

\usepackage[english]{babel}
\usepackage[titles]{tocloft}

\usepackage{amsmath}
\usepackage{graphicx}
\usepackage{algorithm}
\usepackage{algpseudocode}
\usepackage{enumitem}
\usepackage{amsmath}
\usepackage{amssymb}
\usepackage{dsfont}
\usepackage[round]{natbib}
\usepackage[export]{adjustbox}

\usepackage{booktabs}
\usepackage{multirow}

\usepackage{hyperref}
\hypersetup{
    colorlinks=true,
    allcolors=blue,
    pdftitle={Methods of Selective Inference for Linear Mixed Models: a Review and Empirical Comparison},
    pdfpagemode=FullScreen,
}

\title{Methods of Selective Inference for Linear Mixed Models: a Review and Empirical Comparison}
\author{Matteo D'Alessandro$^{*}$, Magne Thoresen}
\date{{\small Oslo Centre for Biostatistics and Epidemiology, Department of Biostatistics, University of Oslo\\%
    $^*$\textit{matteo.dalessandro@medisin.uio.no}\\[2ex]%
    \today}}

\begin{document}

\maketitle

\begin{abstract}

Selective inference aims at providing valid inference after a data-driven selection of models or hypotheses. It is essential to avoid overconfident results and replicability issues. While significant advances have been made in this area for standard regression models, relatively little attention has been given to linear mixed models (LMMs), which are widely used for analyzing clustered or longitudinal data. This paper reviews the existing selective inference approaches developed for LMMs, focusing on selection of fixed effects, where the random effects structure is given. We present these methods in detail and, through comparative simulations, assess their practical performance and computational feasibility under varying data structures. In addition, we apply them to a real-world biological dataset to examine how method choice can impact inference in practice.
Our findings highlight an existing trade-off between computational complexity and statistical power and emphasize the scarcity of methods that perform well as the number of variables increases. In such scenarios, basic sample splitting emerges as the most reliable approach.\\

\end{abstract}

    \section{Introduction}

Classical statistical inferential theory relies on the assumption that all decisions regarding which model to utilize, parameters to estimate, or hypotheses to test are fixed prior to any form of data exploration. More often than not, this goes against common practice in real scientific studies, where analysts are used to selecting the model or the hypotheses through data-driven techniques. This can come through the use of variable selection methods (stepwise selection, AIC, BIC, lasso, \dots), as well as more informal techniques based on data visualization or regression diagnostics \citep{Berk2013}.
Once variable selection has been performed, the properties for inference procedures established by classical theory may no longer hold \citep{Ptscher1991}. As a result, this could lead to invalid inference, thus nullifying the claimed error rates or interpretations. The lack of accounting for explicit or implicit model selection procedures is therefore a contributing factor to the failure of scientific replicability \citep{Benjamini_2009}.

The need for methods that perform valid inference by taking into account the selection step has been emphasized by many authors \citep{Leeb2003, Hjort2003, Benjamini2005}, and different approaches have been explored in the literature under the umbrella term of \textit{selective} (or \textit{post-selection}) \textit{inference}.

Sample splitting \citep{COX1975} is perhaps the easiest approach to inference after selection. The data is split into two parts, a training set and a test set: the first is used to explore the data and select a model, while the second part is used to perform inference on the model selected. Given the independence of observations in the training and test sets, the selected model is not dependent on the data used for inference, and thus the results of classical inference theory still hold.
\cite{Berk2013} establishes the Post-Selection Inference (or short \textit{PoSI}) framework, which constructs general confidence regions that provide \textit{simultaneous inference} for all possible submodels, thus valid under any model selection procedure. Given the generality of the approach, the inference is usually very conservative.
Another possible approach is to consider inference conditional on the selected model: methods in this class have been developed for specific selection procedures such as lasso \citep{Lee2016}, forward stepwise regression and least angle regression \citep{Tibshirani2016}. These approaches are based on mathematically characterizing the selection event, and obtaining the conditional sampling distribution for the estimator in order to provide inference.\\
The methods considered so far aim to cover a regression parameter that depends on the selected model, and is not required to recover the ``true" full model parameters: in fact, there is no assumption of correctness of the full model, which is just interpreted as ``the repository of available predictors" \citep{Berk2013}. A different paradigm is considered in works such as \cite{vandeGeer2014}, which provides inference in high-dimensional regression through the use of the \textit{de-biased} lasso. In this case, the target is the ``true" parameter, and the coverage probability holds unconditionally to the model selected.\\

Linear mixed models (LMMs) are largely used in statistical analysis of clustered or longitudinal data \citep{Laird_1982}, with applications ranging from genome-wide association studies \citep{Jiang2021}, to small area survey estimation \citep{Sugasawa2020} and many more.
Model selection for LMMs can involve both fixed and random effects. Most of the methods available in the literature are adaptations of selection approaches that previously emerged for linear regression models, such as shrinkage methods like lasso \citep{Schelldorfer2014} or SCAD \citep{Ghosh2017}, or use of information criteria (AIC, \cite{Vaida2005}, BIC, \cite{jones2011bayesian}). In this paper we focus on selection and inference for fixed effects exclusively, and will not consider the selection of the random effect structure.

Few examples of methods for inference after selection in LMMs have appeared in the literature so far. \cite{Claesksen2021} considers inference on fixed and mixed parameters (consisting of a linear combination of fixed and random effects), after using a conditional Akaike information criterion (cAIC) as model selection procedure. The authors determine the quadratic constraints that define post-cAIC selection regions, and obtain inference by computing the empirical conditional distribution using Monte Carlo samples.
A similar conditional approach is taken in \cite{Rgamer2022}, where inference is computed for fixed and random effects in the class of additive and linear mixed models, both low and high-dimensional. The approach is very general and allows for computation of selective inference for any selection rule which can be stated as a deterministic function of the response. P-values can once again be approximated numerically through Monte Carlo sampling.\\
The previous two methods both take into consideration conditional inference, and aim to cover a parameter dependent on the selection. The one presented in \cite{Kramlinger2023}, on the other hand, follows an unconditional inference paradigm, with the aim to cover
the true parameter. The paper presents an extension to LMMs of the results in \cite{Ewald2018}, obtaining confidence sets based on the lasso estimator for the whole fixed effect parameter vector that guarantee nominal coverage uniformly over the space of coefficients and covariance parameters. The method's results are only valid for low-dimensional fixed effects, because of the need for consistent estimation of the covariance parameters.\\

The purpose of this paper is a review of the existing approaches for inference after selection of fixed effects in linear mixed models and a comparison of their results in a common simulation study setting. In particular, we will compare the basic sample splitting technique with the methods presented in \cite{Claesksen2021}, \cite{Rgamer2022} and \cite{Kramlinger2023}.

The rest of the paper is structured as follows. In Section \ref{section_method} we introduce the framework for linear mixed models and we describe the considered methods in detail. In Section \ref{section_simulation_studies} we present comparative simulations to evaluate the performance of the methods in different settings of data generation. We provide a comparison on an example biological dataset in Section \ref{section_real_data_example}. We conclude with a discussion of the results in Section \ref{section_discussion}.

    \section{Methods}\label{section_method}

We start by presenting a general notation for the LMM, introducing traditional methods of estimation and inference for its parameters, as well as possible approaches to model selection. We denote by $N$ the total number of clusters, by $n_i$ the number of observations in cluster $i$ and by $n = \sum_{i=1}^N n_i$ the total number of observations. We consider, for each cluster $i$,

\begin{equation} \label{eq:lmm_cluster}
    \mathbf{y}_i=\mathbf{X}_i \boldsymbol{\beta}_0+\mathbf{Z}_i \mathbf{b}_i+\boldsymbol{\varepsilon}_i, \quad i \in\{1, \ldots, N\}, 
\end{equation}
\\
with observations $\mathbf{y}_i \in \mathbb{R}^{n_i}$, covariates $\mathbf{X}_i \in \mathbb{R}^{n_i \times p}, \mathbf{Z}_i \in \mathbb{R}^{n_i \times q}$, $\boldsymbol{\beta}_0 \in \mathbb{R}^p$ the vector of fixed effects, $\mathbf{b}_i \in \mathbb{R}^q$ the vector of random effects, and $\boldsymbol{\varepsilon}_i \in \mathbb{R}^{n_i}$ a vector of random errors.
Furthermore, we assume that $\mathbf{b}_i \sim \mathcal{N}_q (\mathbf{0}_q, \boldsymbol{G}_q(\boldsymbol{\theta}))$ and $\boldsymbol{\varepsilon}_i \sim \mathcal{N}_{n_i} (\mathbf{0}_{n_i}, \boldsymbol{R}_{n_i}(\boldsymbol{\theta}))$ are independent random variables for all clusters $i \in\{1, \ldots, N\}$. Here, $\boldsymbol{\theta} \in \mathbb{R}^{h}$ is a vector of unknown variance parameters. Note that fixed effect coefficients $\boldsymbol{\beta}_0$ are common for all observations,  while the realization of random effects $\mathbf{b}_i$ is different for each cluster: the presence of these coefficients allows the LMM to model both the variance between and within clusters of observations. For example, in a biological study on multiple patients, random effects can be linked to features specific to each subject.\\
We can rewrite Equation (\ref{eq:lmm_cluster}) for all clusters as,

\begin{equation} \label{eq:lmm_tot}
    \mathbf{y}=\mathbf{X} \boldsymbol{\beta}_0+\mathbf{Z} \mathbf{b}+\boldsymbol{\varepsilon}, 
\end{equation}
\\
where $\mathbf{y} = (\mathbf{y}_1, \dots, \mathbf{y}_N)^{\top}$, $\mathbf{X} = (\mathbf{X}_1^{\top}, \dots, \mathbf{X}_N^{\top})^{\top}$, $\mathbf{Z} = \text{diag}(\mathbf{Z}_1, \dots, \mathbf{Z}_N)$ a block-diagonal matrix, $\mathbf{b} = (\mathbf{b}_1, \dots, \mathbf{b}_N)^{\top}$, $\boldsymbol{\varepsilon} = (\boldsymbol{\varepsilon}_1, \dots, \boldsymbol{\varepsilon}_N)^{\top}$. In particular, $\mathbf{X} \in \mathbb{R}^{n \times p}$ and $ \mathbf{Z} \in \mathbb{R}^{n \times Nq}$. We can furthermore define the full variance-covariance matrix for the random effects as $\mathbf{G} = \text{diag}_N(\mathbf{G}_q)$, as well as for the random errors $\mathbf{R} = \text{diag}(\mathbf{R}_{n_1}, \dots \mathbf{R}_{n_N})$. The study of LMMs can be conducted from two distributional viewpoints, by either considering the marginal or conditional distributions of $\mathbf{y}$. The former can be expressed as $\mathbf{y} \sim \mathcal{N}_n(\mathbf{X} \boldsymbol{\beta}_0, \boldsymbol{\Sigma}(\boldsymbol{\theta}))$, with $\Sigma(\boldsymbol{\theta}) = \mathbf{R}(\boldsymbol{\theta}) + \mathbf{Z}\mathbf{G}(\boldsymbol{\theta})\mathbf{Z}^{\top}$, while the latter as $\mathbf{y}|\mathbf{b} \sim \mathcal{N}_n(\mathbf{X} \boldsymbol{\beta}_0 + \mathbf{Z} \mathbf{b}, \mathbf{R}(\boldsymbol{\theta}))$.

When the interest lies primarily on the fixed effects, and we go on to estimate the marginal model, $\boldsymbol{\theta}$ can be estimated through REML (Restricted Maximum Likelihood), computed by EM or Newton-Raphson algorithm \citep{Lindstrom1988}. The estimate $\hat{\boldsymbol{\theta}}$ can then be plugged in to obtain $$\hat{\boldsymbol{\beta}_0} = (\mathbf{X}^{\top}\boldsymbol{\Sigma}(\hat{\boldsymbol{\theta}})^{-1}\mathbf{X})^{-1}\mathbf{X}^{\top}\boldsymbol{\Sigma}(\hat{\boldsymbol{\theta}})^{-1}\mathbf{y}.$$
Inference on the fixed effects can then be obtained through Wald Z-tests \citep{Wald1943} or likelihood ratio tests. Finally, model selection for LMMs, both of fixed and random effects, has been largely explored in the literature. \cite{Vaida2005} propose the conditional AIC, an extension of the AIC, for mixed effects models with a focus on cluster parameters, \cite{Bondell2010} introduce a joint variable selection method for fixed and random effects using a modified Cholesky decomposition. Maximum penalized likelihood methods of selection and estimation have also been proposed, such as SCAD or lasso penalty \citep{Ibrahim2010}.

\subsection{Post-selection inference target} \label{subsection: inference target}

As highlighted in \cite{Berk2013}, two perspectives can be considered when working with parameters of a submodel. In the first, the selection of a submodel is understood as constraining the unselected parameters to be zero, and estimating the selected ones accordingly. In this case, the pool of parameters that are being considered is still the one of the full model, which is given the special status of containing all causal predictors for the response. The second perspective, on the other hand, would argue that the submodel has its own pool of parameters, and the ones that are not selected from the full model are not only constrained to zero, but simply do not exist. The first case corresponds to so called population-based inference targets \citep{Zhang2022}, which are simply the full model coefficients $\boldsymbol{\beta}_{0,1}, \dots, \boldsymbol{\beta}_{0,p}$. The second case aims at covering the non-standard projection-based target which, given the sub-matrix of selected predictors $\mathbf{X}_M$ and assuming known variance-covariance matrix $\boldsymbol{\Sigma}$, can be defined in our case as

\begin{equation} \label{projection_parameter}
\boldsymbol{\beta}_M = \text{arg}\min_{\boldsymbol{\beta} \in \mathbb{R}^{|M|}} ||\mathbb{E}(\mathbf{y}) - \mathbf{X}_M\boldsymbol{\beta}||^2 = (\mathbf{X}_M^{\top}\boldsymbol{\Sigma}^{-1}\mathbf{X}_M)^{-1}\mathbf{X}_M^{\top}\boldsymbol{\Sigma}^{-1}\mathbb{E}(\mathbf{y})
\end{equation}
\\
Observe in particular that, when the first-order correctness of the selected model holds so that $\mathbb{E}(\mathbf{y}) = \mathbf{X}_M\boldsymbol{\beta}_0$, the two described targets coincide.\\
The population-based target is embraced by the work of \cite{Ewald2018} for low-dimensional linear regression, where the lasso estimator is used as base for inference for the parameters of the full model. The same approach is extended to LMMs by \cite{Kramlinger2023}. In their work, although the estimation through lasso forces some coefficients to zero, the construction of confidence sets still aims at covering all true coefficients of the initial full model.
The work by \cite{Rgamer2022}, on the other hand, has a projection-based target as aim of inference. The method doesn't perform inference on the parameters that are not selected, because they are considered non-existent. \cite{Berk2013} argues that this perspective presents some advantages, as it doesn't associate any data-generating property to the predictors in the full model. This assumption in fact, could be hard to implement in practice, as in applications it is usually impossible to include all pertinent predictors for the modeling of the response.

\subsection{Naïve inference}
We first discuss the option of ignoring the effects of selection when performing inference. As shown in a series of works by Hannes Leeb and Benedikt M. Pötscher \citep{Leeb2003, Leeb2006, Leeb2007}, the finite sample distributions of post-selection parameter estimators generally deviate from Gaussian, and have complicated forms obtained through mixtures that depend on the unknown parameter $\beta$. Furthermore, they go on to prove that it is impossible to estimate the post-selection distribution of parameter estimators in a uniform and consistent way. Similar issues arise in the LMM setting, where ignoring the effects of selection can lead to distorted sampling distributions and unreliable confidence intervals. As shown in \cite{Claesksen2021}, failing to account for selection can result in empirical coverage probabilities below the nominal level, particularly for noise variables. These challenges underscore the necessity of using inference methods specifically designed for post-selection settings, which we introduce in the following sections.

\subsection{Sample splitting}

The main idea behind sample splitting stems from the observation that using the same data both for exploration and statistical inference can lead to undesirable distributional results. To prevent this, the observations are split into two parts randomly: a portion of the data is used for model selection (training data) while the remaining portion is used to perform inference based on that selected submodel (test data) \citep{COX1975}. The main advantages of this procedure are the simplicity of its algorithm, as well as the wide range of selection procedures that can be implemented in its use: provided it uses only the training data, no assumptions on the selection method are needed. An example of sample splitting implementation for LMMs would be to select fixed effects through a lasso penalty, and make inference by fitting a standard LMM to the selected set of covariates. We also note that the validity of the sample splitting procedure relies on the independence between the observations in the training and test sets: in the case of clustered data, this implies that the split would have to be performed on the clusters, rather than on the single observations. All samples from the same cluster would then belong to either of the two sets, ensuring independence.

Sample splitting also presents multiple disadvantages. First, the split implies reduced sample size for inference, which results in lower statistical power. Research aimed at reducing the amount of information lost in the split, and thus maximize power, has been carried out in the context of linear models by \cite{Fithian2014}, who proposes a \textit{data carving} approach, and \cite{Tian2018} who produce a randomized response variation on sample splitting. Another issue lays in the sensitivity of the procedure to the random data split, which can produce very variable results both in regards to the selection and the inference. The randomness in the selection also causes trouble with interpretation, since a change in the selection causes a change in the projection-based target of estimation. The aggregation of inference across multiple data splits \citep{Meinshausen_2009}, which could serve as a tool for variability reduction, requires first order correctness of the selected model to produce unbiased p-values for the coefficients in the full model.

\subsection{\cite{Claesksen2021}}

The approach presented in \cite{Claesksen2021} aims at constructing confidence intervals for fixed effects, their linear combinations, and mixed effects (linear combinations that include random effects) in low-dimensional linear mixed models. This method applies when only the fixed effects are selected, while the random structure is given. The authors build on the results of \cite{Charkhi2018}, which are generally applicable to any likelihood-based model, but did not take random effects into account specifically. In particular, the use of the classical AIC in LMMs is based on the use of a marginal likelihood, and therefore cannot be employed if there is interest in prediction of cluster-level effects. For this reason, in order to obtain inference statements on mixed parameters as well, the authors consider a different model selector, the conditional AIC (cAIC), in the formulation of \cite{Kubokawa2011}, which takes into account the estimation of variance parameters.
Selective inference is computed in this case by determining explicitly the constraints that define the selection regions of parameters after cAIC, and approximating the limiting distributions for fixed and mixed parameters conditionally on this regions using Monte Carlo samples. A crucial point in the analysis is the definition of the set of possible candidate models, $\mathcal{M}$: the authors consider both the case of nested models, given an order of inclusion of the variables, the general case of all possible submodels, as well as model misspecification, where the true model does not necessarily exist, and the standard inferential target is not available.
The general algorithm is presented in Algorithm \ref{alg:claeskens}, while we refer to the original article for further details.

\begin{algorithm}
\caption{Method from \cite{Claesksen2021}.}\label{alg:claeskens}
\begin{algorithmic}
\Require Set of candidate models $\mathcal{M}$, number of Monte Carlo samples $B$, nominal coverage level $\alpha$. \\
\vspace{0.5cm}
\begin{enumerate}[label=(\roman*)]
    \item Select model $M$ that has the smallest cAIC between the elements of $\mathcal{M}$.
    \item Calculate the quadratic constrains that define the selection region for $M$.
    \item Select $B$ Monte Carlo samples from a truncated, multivariate normal distribution that satisfies the constraints.
\end{enumerate}
\\
\vspace{0.5cm}
\textbf{Result:} Retrieve the relevant quantiles for probabilities $\alpha/2$ and $1-\alpha/2$ from the empirical post-selection distributions and construct confidence intervals for either fixed or mixed parameters.
   
\end{algorithmic}
\end{algorithm}

In order to compare the performance of the methods with the others presented, we also implemented the procedure to obtain p-values for the testing of fixed effects from the obtained post-selection distribution of each coefficient.\\
Although general in its approach, the algorithm is currently only implemented for a nested error regression models (NERM), which can be modeled as

\begin{equation}\label{eq:nerm}
    y_{ij}=\sum_{k=1}^{p}\beta_kx_{ijk}+u_{i}+e_{ij}
\end{equation}

which corresponds to a linear mixed model with fixed effects and a random intercept. Step (iii) in Algorithm \ref{alg:claeskens} involves sampling from a truncated multivariate normal distribution. This process depends on identifying suitable starting points, which are randomly generated from a normal distribution and then evaluated to ensure they satisfy the quadratic constraints. The variance of this normal distribution plays a crucial role in determining the efficiency of finding valid starting points, significantly impacting computation time. Additionally, as the number of predictors increases, the number of constraints grows exponentially, limiting the method's applicability to models with only a small number of predictors.
The method is implemented in the R package \texttt{postcAIC}, which can be installed from the author's GitHub at \url{https://github.com/KatarzynaReluga/postcAIC}.

\subsection{\cite{Rgamer2022}}

The conditional perspective on inference after selection for LMMs has been developed by \cite{Rgamer2022}. The approach obtains inference results for fixed and random effects in the class of additive and linear mixed models, both low and high-dimensional. Unlike most methods in the area of conditional inference, it allows complete generality of the selection procedure, given that it can be expressed as a deterministic function $\mathcal{S}$ of the response. The authors develop inference in the case of selection of both fixed and random effects but, for the comparison at hand, we will restrict to the case of selection of fixed effects only. In particular, we will focus on the marginal distributional viewpoint for LMMs, assuming that $\mathbf{Y} \sim \mathcal{N}(\boldsymbol{\mu}, \boldsymbol{\Sigma}(\boldsymbol{\theta}))$. The use of the selective inference framework allows us to relax the assumption of model correctness by focusing on a projection-based target of inference. In particular, the response mean $\boldsymbol{\mu}$ is allowed to have an arbitrary structure, possibly depending on unobserved covariates. The model equation in (\ref{eq:lmm_tot}) is, in this case, only regarded as a \textit{working model}, with $X$ being the repository of all available predictors. After the selection of fixed effects $\mathcal{S}(\mathbf{y}) = M \subset \{1, \ldots, p\}$, the goal is to infer on $\boldsymbol{\beta}_M$, which represent the coefficients for the projection of $\boldsymbol{\mu}$ onto the column space of $\mathbf{X}_M$.
We now go on to describe the procedure in the case of testing a single coefficient $\beta_{M,j}$. The null hypothesis we are looking to test in this case is $H_0: \beta_{M,j} = \rho$, although the authors discuss the case of testing groups of variables as well. Assuming a known variance-covariance matrix $\boldsymbol{\Sigma}$, the proposed test statistic is $T = \hat{\beta}_{M,j} = \mathbf{v}^{\top}\mathbf{Y}$ with  $\mathbf{v} = \boldsymbol{\Sigma}^{-1}\mathbf{X}_M(\mathbf{X}_M^{\top}\boldsymbol{\Sigma}^{-1}\mathbf{X}_M)^{-1}\mathbf{e}_j$ where $\mathbf{e}_j$ is the standard basis vector, with 1 in the $j$-th position and 0 elsewhere. The test statistic is shown to have null distribution $\mathcal{N}(\rho,\mathbf{e}_j^{\top}(\mathbf{X}_M^{\top}\boldsymbol{\Sigma}^{-1}\mathbf{X}_M)^{-1}\mathbf{e}_j = \kappa)$. We are now interested in performing inference using $T$, conditional on the selection of $M$, thus restricting $T$ to assume values in a set $\mathcal{T}$. The selection event can be described explicitly for some selection rules: in the example of lasso, \cite{Lee2016} obtains the characterization of the selection event as a polyhedron.. In the general case, however, $\mathcal{T}$ cannot be characterized. As a solution to determining the distribution of $T$ restricted to $\mathcal{T}$, the authors utilize a numerical procedure based on Monte Carlo sampling, building on the work of \cite{Yang2016}. P-values for the observed value of the test statistic $t_{obs}$ are calculated through the function

\begin{equation}\label{eq: rugamer_pvalue_function}
f(\rho) = \frac{\mathbb{E}_{T \sim \mathcal{N}(0,\kappa)}[ e^{T \rho / \kappa} \mathds{1}\{ T \in \mathcal{T}, T > t_{obs} \} ]}{\mathbb{E}_{T \sim \mathcal{N}(0,\kappa)}[ e^{ T \rho / \kappa} \mathds{1} \{ T \in \mathcal{T}\} ]}
\end{equation}

where the expectations can be replaced by sample averages. In particular, observing that $Y$ can be decomposed as $
\mathbf{Y}= \frac{\mathbf{v}}{\mathbf{v}^\top \mathbf{v}} T + \mathbf{P}_{\mathbf{v}}^\perp\mathbf{Y}$ with $\mathbf{P}_{\mathbf{v}}^\perp = \mathbf{I}_n - \frac{\mathbf{v}\mathbf{v}^\top}{\mathbf{v}^\top \mathbf{v}}$, the Monte Carlo procedure can be obtained by drawing samples $T^b$ from the normal distribution, defining

\begin{equation}\label{eq:yb}
    \mathbf{Y}^b = \frac{\mathbf{v}}{\mathbf{v}^\top \mathbf{v}} T^b + \mathbf{P}_{\mathbf{v}}^\perp\mathbf{Y}
\end{equation}

and checking the condition $\mathcal{S}(\mathbf{Y}^b) = M$. Furthermore, the hypothesis test can be inverted to construct confidence intervals for the coefficients that take into account the selection. Given a nominal value $\alpha \in \{ 0,1 \}$, the lower and upper bounds of the confidence interval can be found by numerically solving $f(\rho) = \alpha/2$ and $f(\rho) = 1 - \alpha/2$ respectively.
Furthermore, the authors propose the use of an importance sampler to deal with cases in which the numerical approximation of (\ref{eq: rugamer_pvalue_function}) could present issues. The procedure to obtain inference from \cite{Rgamer2022} is described in Algorithm \ref{alg:rugamer}.

\begin{algorithm}
\caption{Method from \cite{Rgamer2022}.}\label{alg:rugamer}
\begin{algorithmic}
\Require Selection rule $\mathcal{S}$, covariance matrix $\boldsymbol{\Sigma}$, number of samples $B$. \\
\vspace{0.5cm}
\begin{enumerate}[label=(\roman*)]
    \item Obtain the selected model $M = \mathcal{S}(\mathbf{y})$.
    \item Obtain the variance for the test statistic $\kappa = \mathbf{e}_j^{\top}(\mathbf{X}_M^{\top}\boldsymbol{\Sigma}^{-1}\mathbf{X}_M)^{-1}\mathbf{e}_j$.
    \item Generate samples $T^b$, $b = 1, \dots, B$ from $\mathcal{N}(0,\kappa)$.
\end{enumerate}

\vspace{0.3cm}

\For{$b \in \{1, \dots, B\}$}\\
\vspace{0.1cm}
\begin{enumerate}
\item Define $\mathbf{Y}^b$ as in (\ref{eq:yb}).
\item Check if $T^b \in \mathcal{T}$ by verifying if $\mathcal{S}(\mathbf{Y}^b) = M$.
\end{enumerate}
\vspace{0.1cm}
\EndFor
\vspace{0.2cm}
\\
Approximate function $f$ in (\ref{eq: rugamer_pvalue_function}) by the empirical average obtained from the samples.
\vspace{0.2cm}
\\
\textbf{Result:} the p-value for testing the null hypothesis $H_0: \beta_{M,j} = 0$ can be obtained as $f(0)$.
The two-sided confidence interval for $\beta_{M,j}$ at nominal level $\alpha$ is $[L,U]$ with $L$, $U$ the unique solutions of $f(L)=\alpha/2$, $f(U)=1-\alpha/2$.

\end{algorithmic}
\end{algorithm}

One of the key assumptions behind the method is the knowledge of the error covariance structure $\boldsymbol{\Sigma}$. Since this assumption is not realistic in practice, different estimating matrices for $\boldsymbol{\Sigma}$ are explored: possible choices are the estimated covariance of the chosen model, or the more conservative option of the variance-covariance estimator of the intercept model with the given random effects structure, which excludes all fixed effects. Given the need to obtain the conditional inference through a numerical approximation, the method can present a significant computational burden, as the selection has to run $B$ times to check if $T^b \in \mathcal{T}$ for each sample. Additionally, the stability of the selection process plays a crucial role. Only the samples that fall inside the selection event are used for inference, meaning that if the selection is not stable (i.e., small variations in $y$ lead to different selections), the sampling process becomes increasingly difficult. As a result, a large number of samples is required to maintain sufficient power for inference. This issue becomes particularly relevant as the number of predictors grows, since many selection methods, such as lasso, exhibit instability in high-dimensional settings \citep{Xu2012}. The method is implemented in the R package \texttt{selfmade} (SELective inference For Mixed and ADditive model Estimators), which can be installed from the CRAN archive.

\subsection{\cite{Kramlinger2023}}

Assuming a model of the form (\ref{eq:lmm_tot}), the method introduced by \cite{Kramlinger2023} aims at constructing confidence regions based on the the lasso estimator for the fixed effects vector $\boldsymbol{\beta}_0$  which are uniformly valid over the space of coefficients and covariance parameters. This is achieved as an adaptation of the work in \cite{Ewald2018}, which obtains the same result for the standard linear model. In order to justify this extension, the authors prove a uniform consistency result for the REML estimators for the covariance parameters $\boldsymbol{\theta}$, that allows for the construction of confidence sets that hold coverage over all possible values of $\boldsymbol{\theta}$.
After obtaining the REML estimator $\hat{\boldsymbol{\theta}}$ for $\boldsymbol{\theta}$, the lasso estimator is defined as:

\begin{equation}\label{eq:lasso_estim}
\hat{\boldsymbol{\beta}}_L=\underset{\boldsymbol{\beta} \in \mathbb{R}^p}{\operatorname{argmin}}\{\|\mathbf{\Sigma}(\hat{\boldsymbol{\theta}})^{-1 / 2}(\mathbf{y}-\mathbf{X} \boldsymbol{\beta})\|^2+2 \sum_{j=1}^p \lambda_j|\beta_j\}
\end{equation}

The method aims at obtaining confidence regions for $\boldsymbol{\beta}_0$ based on $\hat{\boldsymbol{\beta}}_L$. Given a nominal coverage level $\alpha \in (0,1)$, this can be formally expressed as finding a set $M \in \mathbb{R}^p$ such that $\inf _{\beta_0, \theta} \operatorname{Pr}_{\beta_0, \theta}\{\sqrt{n}(\hat{\beta}_L-\beta_0) \in M\} \approx 1-\alpha$. The approach is based on the observation that the infimum coverage probability will be obtained by considering the cases in which the components of $\boldsymbol{\beta}$ go to infinity in absolute value. Thus, if we construct the $2^p$ minimizers $\hat{\mathbf{u}}_{\mathbf{d}}$  of (\ref{eq:lasso_estim}) associated to all possible combinations of signs $\mathbf{d} \in\{-1,1\}^p$ of the coefficients in $\boldsymbol{\beta}$, $M$ can be chosen as an ellipsoid that contains each of these minimizers with a probability of at least $1-\alpha$. Furthermore, through distributional results in \cite{Ewald2018}, the minimization over $\mathbf{d} \in\{-1,1\}^p$ can be shifted to the choice of the non-centrality parameter for a $\chi^2$-distribution. The procedure for the method is presented in Algorithm \ref{algorithm:kramlinger}. We note that, since the algorithm has to obtain the maximum over a set of $2^p$ elements, its complexity is exponential in the number of predictors in the full model, which makes it unfeasible for models with an increasing $p$.

\begin{algorithm}
\caption{Method from \cite{Kramlinger2023}.}\label{alg:kramlinger}
\begin{algorithmic}
\Require $\lambda_1, \dots, \lambda_p$ shrinkage parameters, $\alpha$ the nominal coverage level, $\hat{\boldsymbol{\theta}}$ the REML estimator for the covariance parameters\\
\vspace{0.5cm}
\begin{enumerate}[label=(\roman*)]
    \item Obtain the lasso estimator $\hat{\boldsymbol{\beta}}_L$ as in (\ref{eq:lasso_estim}).
    \item Define $\hat{\mathbf{C}}=n^{-1} \mathbf{X}^{\top} \mathbf{\Sigma}(\hat{\boldsymbol{\theta}})^{-1} \mathbf{X}$
    \item Define $\boldsymbol{\Lambda}=n^{-1 / 2} \operatorname{diag}\left(\lambda_1, \ldots, \lambda_p\right)$
    \item Obtain $\hat{\tau}$ as 
    $$\hat{\tau}=\max _{\mathbf{d} \in\{-1,1\}^p} \chi_{p, 1-\alpha}^2(\|\hat{\mathbf{C}}^{-1 / 2} \boldsymbol{\Lambda} \mathbf{d}\|^2)$$
\end{enumerate}
\\

\textbf{Result:} The confidence set 

$$M=\{\boldsymbol{\beta} \in \mathbb{R}^p: n\|\hat{\mathbf{C}}^{1 / 2}(\hat{\boldsymbol{\beta}}_L-\boldsymbol{\beta})\|^2 \leq \hat{\tau}\}$$
   
\end{algorithmic}
\label{algorithm:kramlinger}
\end{algorithm}

We adapt this method from its original formulation, which constructs confidence regions for the entire parameter vector \( \boldsymbol{\beta}_0 \), to instead provide confidence intervals and p-values for individual coefficients, as we will compare methods based on the selection of individual variables, evaluating metrics such as power, FWER and FDR.
This adaptation, however, comes at a cost, as performing inference separately for each $\boldsymbol{\beta}_{0j} $ leads to a loss of power. By considering one coefficient at a time rather than the full vector, we sacrifice the ability to leverage dependencies between the variables, resulting in reduced statistical efficiency (the volume of the hypercube of individual confidence intervals will be larger than the volume of the ellipsoid).

When looking for inference on coefficient $\boldsymbol{\beta}_{0j}$ we can obtain the general non-centrality parameter as 

\begin{equation}
\delta = \max _{\mathbf{d} \in\{-1,1\}^p} \|\hat{\mathbf{C}}^{-1 / 2} \boldsymbol{\Lambda} \mathbf{d}\|^2
\end{equation}

and the resulting confidence interval as

\begin{equation}\label{eq:kramlinger_confidence_interval}
\Bigl(\hat{\boldsymbol{\beta}}_{Lj} - \sqrt{\chi_{1, 1-\alpha}^2(\delta) \mathbf{\hat{C}^{-1}}_{j,j}/n},  \quad
    \hat{\boldsymbol{\beta}}_{Lj} + \sqrt{\chi_{1, 1-\alpha}^2(\delta) \mathbf{\hat{C}^{-1}}_{j,j}/n}\Bigr).
\end{equation}

The method relies on regularity assumptions that allow the uniformly consistent estimation of the fixed effects $\boldsymbol{\beta}$ and covariance parameters $\boldsymbol{\theta}$ through REML. In particular, it is only valid for a low-dimensional model ($p<n$). It could be argued that in such a setting, inference on $\boldsymbol{\beta}_0$ could be obtained without the need for selection through lasso: however, as motivated by the authors of \cite{Ewald2018}, the lasso estimator is still used in low-dimensional frameworks, thus entailing the need for methods to obtain valid inference.


\section{Simulation studies}\label{section_simulation_studies}

We go on to study the described methods in simulations that aim to investigate their performance in a range of data settings.
The comparison is made between: \cite{Claesksen2021} referred to as postcAIC, \cite{Rgamer2022} referred to as selfmade, \cite{Kramlinger2023} referred to as UVILassoLMM (Uniformly Valid Inference for LASSO in LMM), as well as naïve inference, to give an insight into the consequence of ignoring selection, and data splitting as a general benchmark. The methods considered consist of a variable selection step followed by an inference step. Naïve inference, data splitting, and selfmade with lasso are implemented using the R package \texttt{glmmlasso}, \cite{Groll2012}, which accounts for random effects during selection. Here, the penalization parameter $\lambda$ is selected through BIC; a more thorough investigation of the impact of the choice of $\lambda$ is carried out in Section \ref{section_lambda_selection}. UVILassoLMM also relies on lasso but estimates the covariance structure separately before applying traditional lasso. The selfmade method, which is generally applicable to any selection procedure, is also considered with backward stepwise selection (selfmade-step), as in the original paper. Finally, postcAIC is designed for inference after selection using the conditional AIC criterion.

The simulated data is generated from model (\ref{eq:lmm_tot}). Given the number of clusters $N$, the number of observations per cluster $n_i$, the number of covariates $p$, we generate the design matrix $\mathbf{X}$ as a multivariate normal $\mathcal{N}_p(0, \boldsymbol{\Sigma}_{\mathbf{X}})$ with $\boldsymbol{\Sigma}_{\mathbf{X}, j,k}$ equal to $1$ if $j=k$ and $0.3$ otherwise. Given the number of non-zero coefficients, the vector $\boldsymbol{\beta}_0$ is obtained by sampling from $\{-2,-1,1,2\}$ with uniform probability for the signal variables and $0$ for the noise variables. The random noise is generated as i.i.d. normal with variance computed to obtain a SNR (signal to noise ratio) of $2$. Unless otherwise specified, we consider just a random intercept with variance equal to that of the random noise.\\

\begin{figure}
    \centering
    \includegraphics[width=1\linewidth, center]{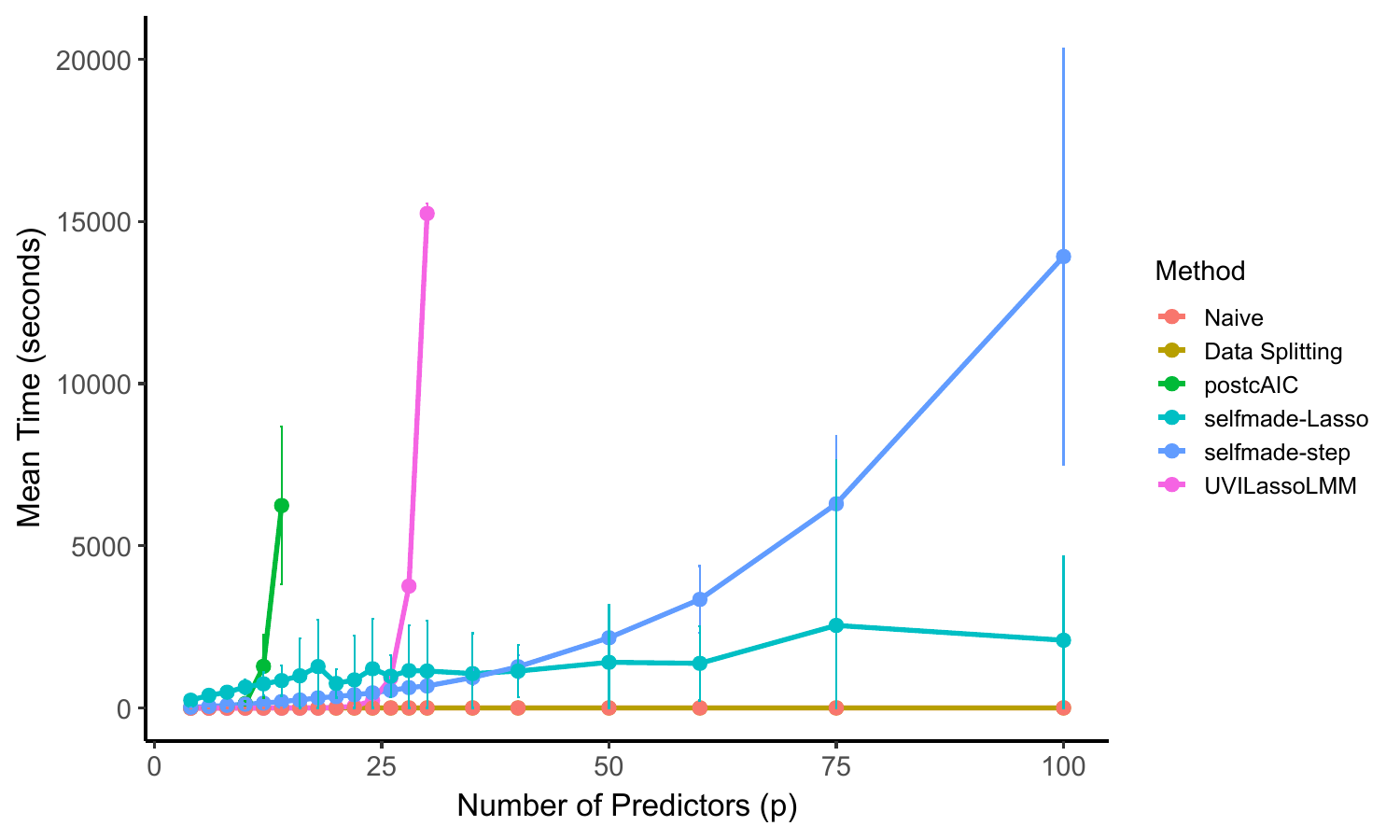}

    \caption{\textit{Comparison of mean computation times (in seconds) across different methods as a function of the number of predictors ($p$). Error bars represent one standard deviation above and below the mean. The plot illustrates the scalability of each method with increasing predictors, highlighting significant differences in computational efficiency.}}
    \label{fig:simulation_time}
\end{figure}

We first compare computational scalability. Figure \ref{fig:simulation_time} illustrates the computational time required by different inference methods as the number of predictors increases up to 100 while keeping the number of observations fixed at $200$, with $40$ subjects each having five repeated measurements. We repeat the simulation $100$ times and report at average time elapsed for each repetition. The results show that postcAIC exhibits a very fast growth in computational time, making it impractical beyond $p=15$. UVILassoLMM follows a similar exponential increase and was discontinued after $p=30$ due to computational infeasibility. The nature of these methods results in an exponential computational increase. In contrast, both naïve inference and data splitting require negligible computational resources regardless of the number of predictors: they have very similar results and are overlapped in the plot. The selfmade demonstrates varying computational trends based on the selection method employed. When using lasso-based selection, the increase in computation time is very slight, whereas stepwise selection leads to a rapid rise in computational demands after $p=50$.  The computational trends in Figure \ref{fig:simulation_time} align with theoretical expectations, where methods involving complex sampling or model space optimization show exponential growth, while simpler fitting methods remain computationally efficient. For larger models, selfmade-Lasso and data splitting emerge as feasible alternatives, while postcAIC and UVILassoLMM become computationally prohibitive.\\

We go on to compare the inference performance of the methods in Table \ref{table:simulation1}. Each simulation is repeated 500 times and average performance metrics are reported. We compute power (TPR) as the number of rejected null hypotheses that are actually false over the total number of non-zero coefficients. As an error metric we use family-wise error rate (FWER). Finally, we look at coverage, computed as the percentage of confidence intervals that actually cover their respective true parameter averaged over signal and noise coefficients separately, and we report the average length of the confidence intervals produced, together with the standard deviation of this length.
The number of covariates $p$ is fixed at $6$, with $3$ non-zero coefficients. We consider both the case of growing number of clusters ($20$ to $50$) and growing number of observations per cluster ($10$ to $25$). We aim to control FWER  at $5\%$ by correcting p-values with the Bonferroni-Holm procedure \citep{Holm1979}. The results show how, even in a small data example, where the effect of selection is limited, naïve inference consistently violates the set error level and shows reduced coverage for the noise variables. The other methods that account for the selection work well and are able to mantain FWER under the set level, although they suffer a loss of power when compared to naïve inference. Data splitting and UVILassoLMM show the lowest power, especially in the examples with a lower number of observations. In terms of power, postcAIC seems to achieve the best performance among the methods that control FWER. The selfmade-step is the most conservative out of the methods considered, returning FWER results well below the control level of $5\%$.

\begin{table}
\begin{center}
\resizebox{\textwidth}{!}{
\begin{tabular}{@{}lllcccccl@{}}
\toprule
\textbf{N}       & \textbf{$n_i$}          & \textbf{Method} & \textbf{TPR} & \textbf{FWER} & \textbf{Signal Cov} & \textbf{Noise Cov} & \textbf{CI (SD)} &  \\ \midrule
\multirow{6}{*}{20} & \multirow{6}{*}{5}  & Naïve           & 0.927        & 0.092         & 0.941               & 0.824              & 1.073 (0.089)    &  \\
                    &                     & Data Splitting  & 0.604        & 0.018         & 0.939               & 0.936              & 1.586 (0.183)    &  \\
                    &                     & postcAIC       & 0.826        & 0.03          & 0.992               & 0.928              & 1.506 (0.147)    &  \\
                    &                     & selfmade-Lasso   & 0.728        & 0.034         & 0.943               & 0.952              & 1.696 (1.168)    &  \\
                    &                     & selfmade-step      & 0.74         & 0.008         & 0.934               & 0.933              & 1.966 (1.932)    &  \\
                    &                     & UVILassoLMM      & 0.581        & 0.02          & 0.987               & 0.969              & 1.689 (0.46)     &  \\ \midrule
\multirow{6}{*}{40} & \multirow{6}{*}{5}  & Naïve           & 0.995        & 0.11          & 0.944               & 0.854              & 0.789 (0.051)    &  \\
                    &                     & Data Splitting  & 0.891        & 0.024         & 0.944               & 0.948              & 1.133 (0.09)     &  \\
                    &                     & postcAIC       & 0.983        & 0.028         & 0.994               & 0.932              & 1.088 (0.076)    &  \\
                    &                     & selfmade-Lasso   & 0.917        & 0.036         & 0.949               & 0.949              & 1.015 (0.356)    &  \\
                    &                     & selfmade-step      & 0.945        & 0.004         & 0.944               & 0.978              & 1.327 (1.608)    &  \\
                    &                     & UVILassoLMM      & 0.829        & 0.028         & 0.987               & 0.976              & 1.178 (0.252)    &  \\ \midrule
\multirow{6}{*}{50} & \multirow{6}{*}{5}  & Naïve           & 0.999        & 0.1           & 0.929               & 0.891              & 0.693 (0.036)    &  \\
                    &                     & Data Splitting  & 0.947        & 0.046         & 0.939               & 0.936              & 0.988 (0.069)    &  \\
                    &                     & postcAIC       & 0.995        & 0.026         & 0.988               & 0.941              & 0.957 (0.059)    &  \\
                    &                     & selfmade-Lasso   & 0.91         & 0.042         & 0.932               & 0.944              & 0.962 (0.554)    &  \\
                    &                     & selfmade-step      & 0.977        & 0.006         & 0.933               & 0.965              & 0.901 (0.717)    &  \\
                    &                     & UVILassoLMM      & 0.921        & 0.024         & 0.974               & 0.977              & 1.001 (0.219)    &  \\ \midrule
\multirow{6}{*}{10} & \multirow{6}{*}{10} & Naïve           & 0.933        & 0.086         & 0.941               & 0.829              & 1.038 (0.096)    &  \\
                    &                     & Data Splitting  & 0.601        & 0.014         & 0.922               & 0.937              & 1.528 (0.176)    &  \\
                    &                     & postcAIC       & 0.827        & 0.018         & 0.991               & 0.961              & 1.48 (0.149)     &  \\
                    &                     & selfmade-Lasso   & 0.762        & 0.008         & 0.961               & 0.973              & 1.723 (1.22)     &  \\
                    &                     & selfmade-step      & 0.749        & 0.008         & 0.938               & 0.944              & 1.869 (1.781)    &  \\
                    &                     & UVILassoLMM      & 0.583        & 0.012         & 0.983               & 0.983              & 1.652 (0.415)    &  \\ \midrule
\multirow{6}{*}{10} & \multirow{6}{*}{20} & Naïve           & 1            & 0.112         & 0.941               & 0.745              & 0.709 (0.04)     &  \\
                    &                     & Data Splitting  & 0.92         & 0.026         & 0.947               & 0.944              & 1.017 (0.077)    &  \\
                    &                     & postcAIC       & 0.995        & 0.022         & 0.993               & 0.948              & 1.014 (0.062)    &  \\
                    &                     & selfmade-Lasso   & 0.942        & 0.05          & 0.924               & 0.912              & 1.155 (0.688)    &  \\
                    &                     & selfmade-step      & 0.97         & 0.008         & 0.943               & 0.944              & 1.04 (0.913)     &  \\
                    &                     & UVILassoLMM      & 0.897        & 0.022         & 0.981               & 0.972              & 1.047 (0.224)    &  \\ \midrule
\multirow{6}{*}{10} & \multirow{6}{*}{25} & Naïve           & 1            & 0.09          & 0.953               & 0.765              & 0.652 (0.031)    &  \\
                    &                     & Data Splitting  & 0.957        & 0.016         & 0.953               & 0.954              & 0.931 (0.061)    &  \\
                    &                     & postcAIC       & 0.999        & 0.018         & 0.995               & 0.957              & 0.929 (0.05)     &  \\
                    &                     & selfmade-Lasso   & 0.949        & 0.006         & 0.961               & 0.987              & 0.9 (0.437)      &  \\
                    &                     & selfmade-step      & 0.977        & 0             & 0.954               & 0.971              & 0.921 (0.793)    &  \\
                    &                     & UVILassoLMM      & 0.941        & 0.012         & 0.983               & 0.984              & 0.986 (0.2)      &  \\ \bottomrule
\end{tabular}
}
\end{center}
\caption{\textit{Power, FWER, coverage probabilities and average lengths of confidence intervals for the methods considered. FWER is controlled at $5\%$ and nominal coverage probability is $95\%$.}
 }
 \label{table:simulation1}
 
\end{table}

\begin{figure}
    \includegraphics[width=1\linewidth]{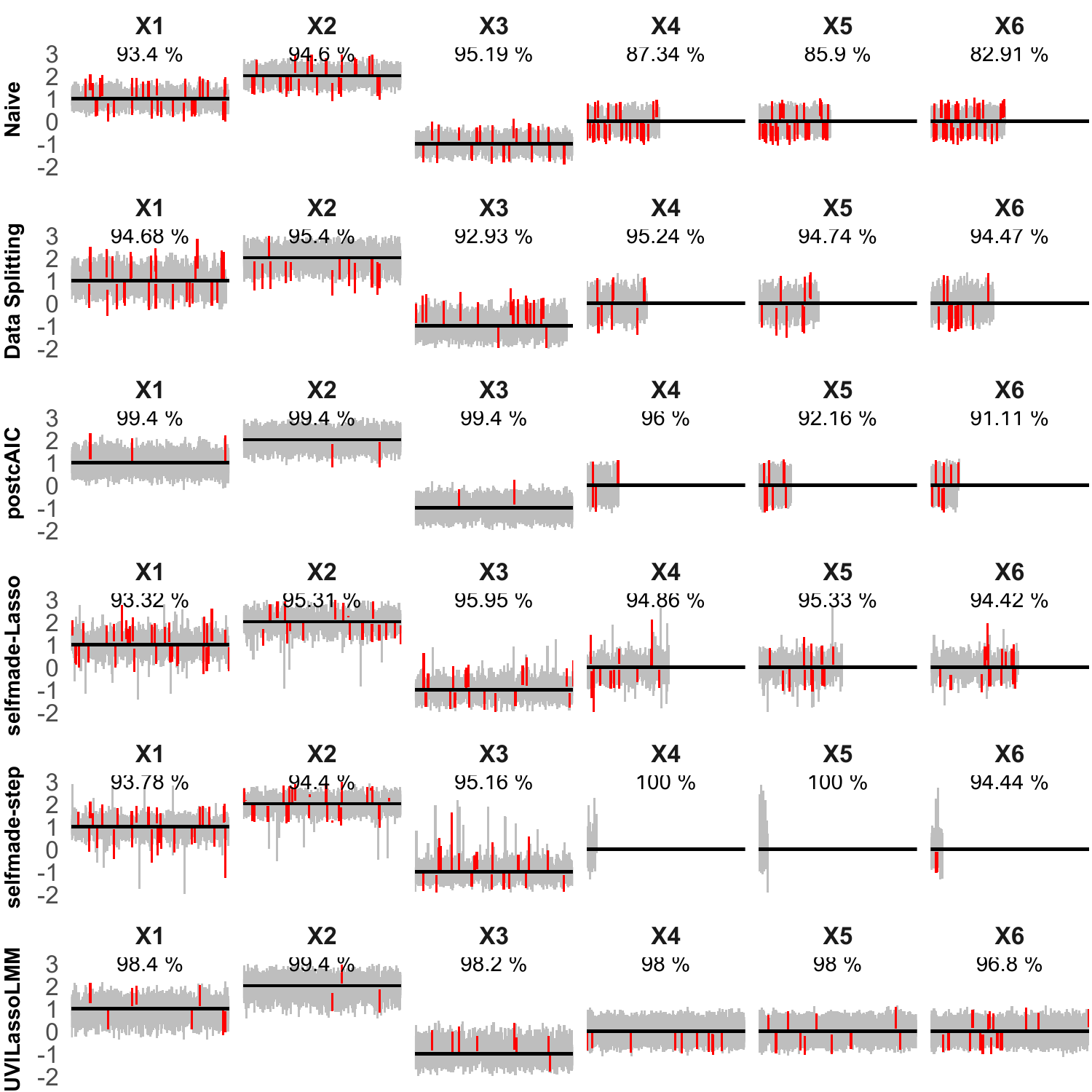}
    \caption{\textit{Computed confidence intervals and their coverage rates for 500 realizations of the simulated model with $N = 40$, $n_i=5$,
$p = 6$. The first three coefficients are non-zero. The grey confidence intervals cover their true coefficient, while the red ones do not. The percentages reported are the empirical coverages for each coefficient. All confidence intervals are on the same scale. For all methods except UVILassoLMM, only the confidence intervals of the variables selected are computed and thus reported. UVILassoLMM computes an interval even for the coefficients estimated as zero after lasso, and thus all intervals are reported for all variables.}}
    \label{fig:confidence_interval_plot}
\end{figure}

Figure \ref{fig:confidence_interval_plot} shows the computed confidence intervals for all coefficients in the simulated model over the 500 iterations. The confidence intervals that do not cover the ground truth coefficients are colored in red. The figure also reports the coverage percentage for each method and coefficient. The Figure shows additional features of the selection performed by the different methods. The selfmade-step and postcAIC are the methods that include the least number of noise variables in the selection. Both conditional approaches of selfmade shows great variability in confidence interval length, with some intervals being much larger. This is a common feature of methods that rely on conditioning on the selection event (see Section 6 of \cite{Lee2016} for more details).

\begin{figure}
    \centering
    \includegraphics[width=1\linewidth, center]{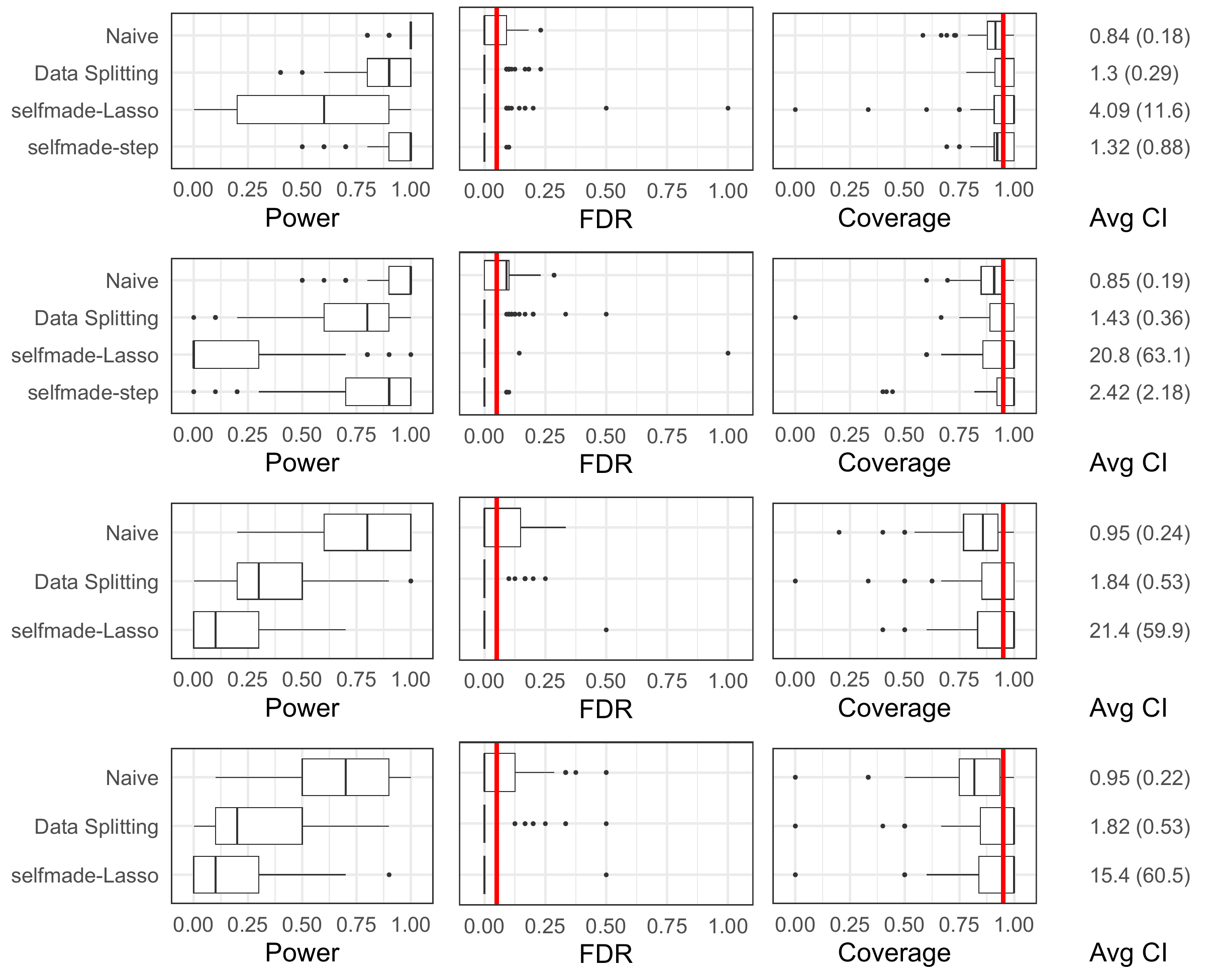}

    \caption{\textit{Power, FDR and coverage probability for different methods of inference after selection in LMMs. The desired levels for FDR and coverage are highlighted by the vertical red lines. All simulations have number of observations $n=200$. From top to bottom, the number of predictors $p$ is respectively $25$, $50$, $150$ and $250$, with $10$ non-zero coefficients.}}
    \label{fig:simulation_large}
\end{figure}

Figure \ref{fig:simulation_large} presents results from the larger simulation setting, where the number of observations remains fixed at 200 while the number of predictors $p$ increases from 25 to 250. In this simulation, where the number of predictors is larger and controlling FWER would risk resulting in a very conservative selection, we aim at controlling the false discovery rate (FDR) at $5\%$ by correcting p-values with the Benjamini-Hockberg procedure \citep{Benjamini1995}. We don't include UVILassoLMM and postcAIC here: as shown in Figure \ref{fig:simulation_time}, the use of these methods becomes computationally unfeasible for more than $30$ parameters.

The results highlight that selfmade-step achieves the highest power among the selective inference methods whenever it is computationally feasible to use. However, due to computational constraints, stepwise selection is excluded for $p=150$ and $p=250$. In contrast, selfmade-Lasso exhibits a steep decline in power as $p$ increases, becoming almost ineffective beyond $p=50$. Additionally, its confidence intervals grow substantially in length. This poor performance of selfmade-Lasso can be attributed to the high selection instability associated with lasso as the number of predictors increases. The selfmade relies on a Monte Carlo approximation, where only the samples that return the same selection of predictors as the original response are used for inference. If the selection process is highly unstable, as is often the case with lasso when the number of noise variables increases, the number of valid samples available for inference becomes very low. This results in a significant loss of power for the method. In this setting with $p=25$, on average $93\%$ of samples yield the same selection for selfmade-step, while for selfmade-Lasso it is only $13\%$. To make effective use of selfmade in larger settings, we would either need to consistently increase the computational time to compute more samples, or alternatively, employ a selection method that is able to scale computationally while still providing stable selection across varying levels of $p$.

\subsection{Increasing random intercept variance}

\begin{figure}
    \centering
    \includegraphics[width=1\linewidth, center]{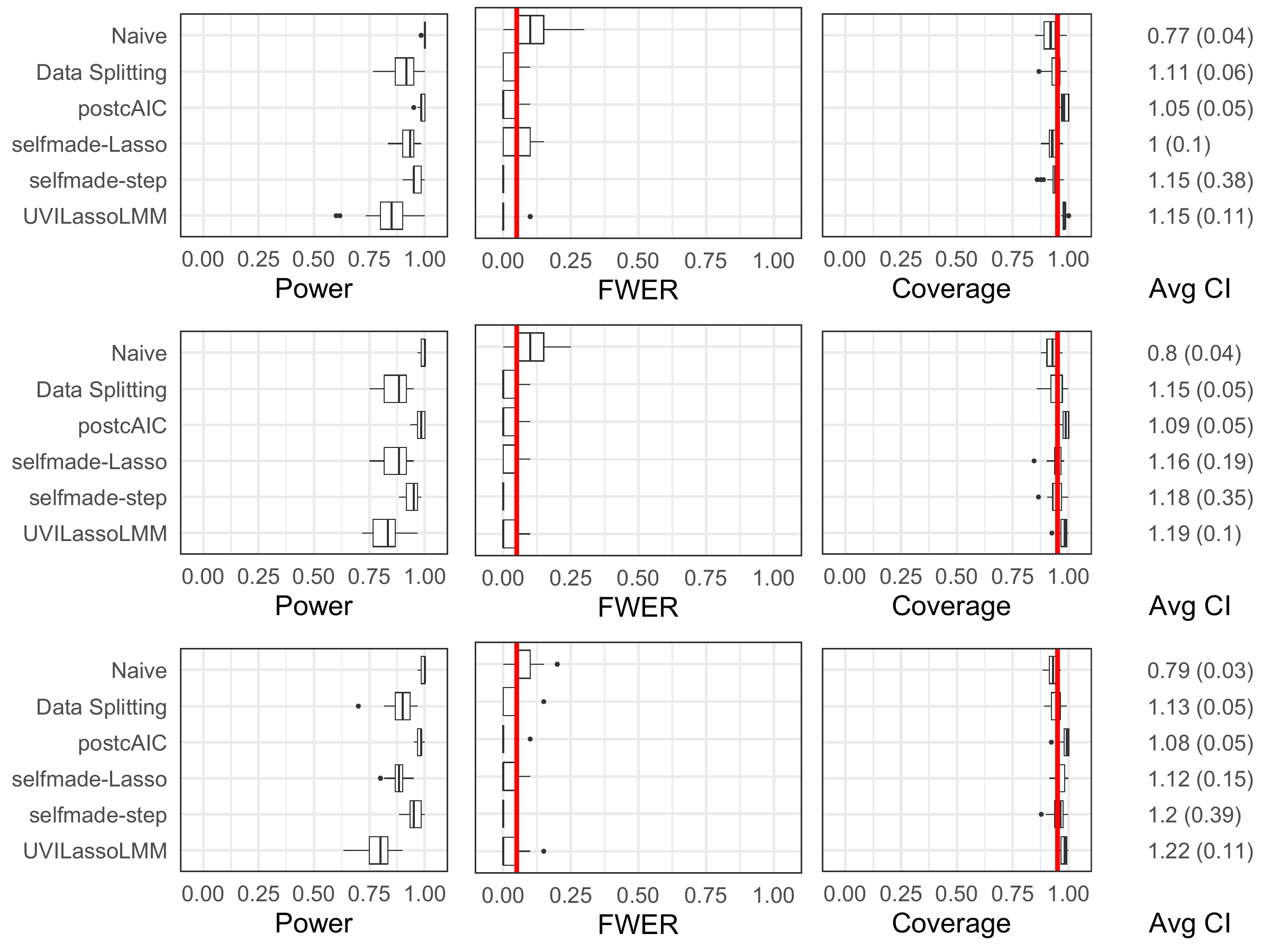}

    \caption{\textit{Power, FWER and coverage probability for different methods of inference after selection in LMMs. The desired levels for FWER and coverage are highlighted by the vertical red lines. All simulations have $n=200$, $p=6$ and $3$ non-zero coefficients. From top to bottom, the variance of the random intercept is respectively $3$, $6$, $12$.}}
    \label{fig:simulation_changingsigv}
\end{figure}

Figure \ref{fig:simulation_changingsigv} investigates the impact of the random effect structure. Specifically, we consider the case of $N=40$, $n_i=5$, $p=6$ with 3 non-zero coefficients and a random intercept of increasing variance ($\sigma^2 = 3, 6, 12$). Overall, all methods considered maintain stable levels of power, FWER, and coverage for all settings. All selection methods employed (cAIC, glmmLasso, and LMM stepwise selection) are explicitly designed for LMMs and inherently account for the correlation structure induced by the random effects. As a result, the increasing variance of the random intercept does not negatively impact variable selection or inference quality.

\subsection{Choice of $\lambda$}\label{section_lambda_selection}

A central question when using lasso lies in the choice of the penalization parameter $\lambda$. Moreover, the setting of selective inference introduces an additional challenge, as most methods of inference after selection are developed for a value of $\lambda$ that is fixed a priori or selected independently of the data, while in practice this value is usually selected through cross-validation or through the use of an information criterion. The use of the cross-validation procedure in selective inference for the linear model has been explored in \cite{Loftus2015}.

\begin{figure}
    \centering
    \includegraphics[width=1\linewidth, center]{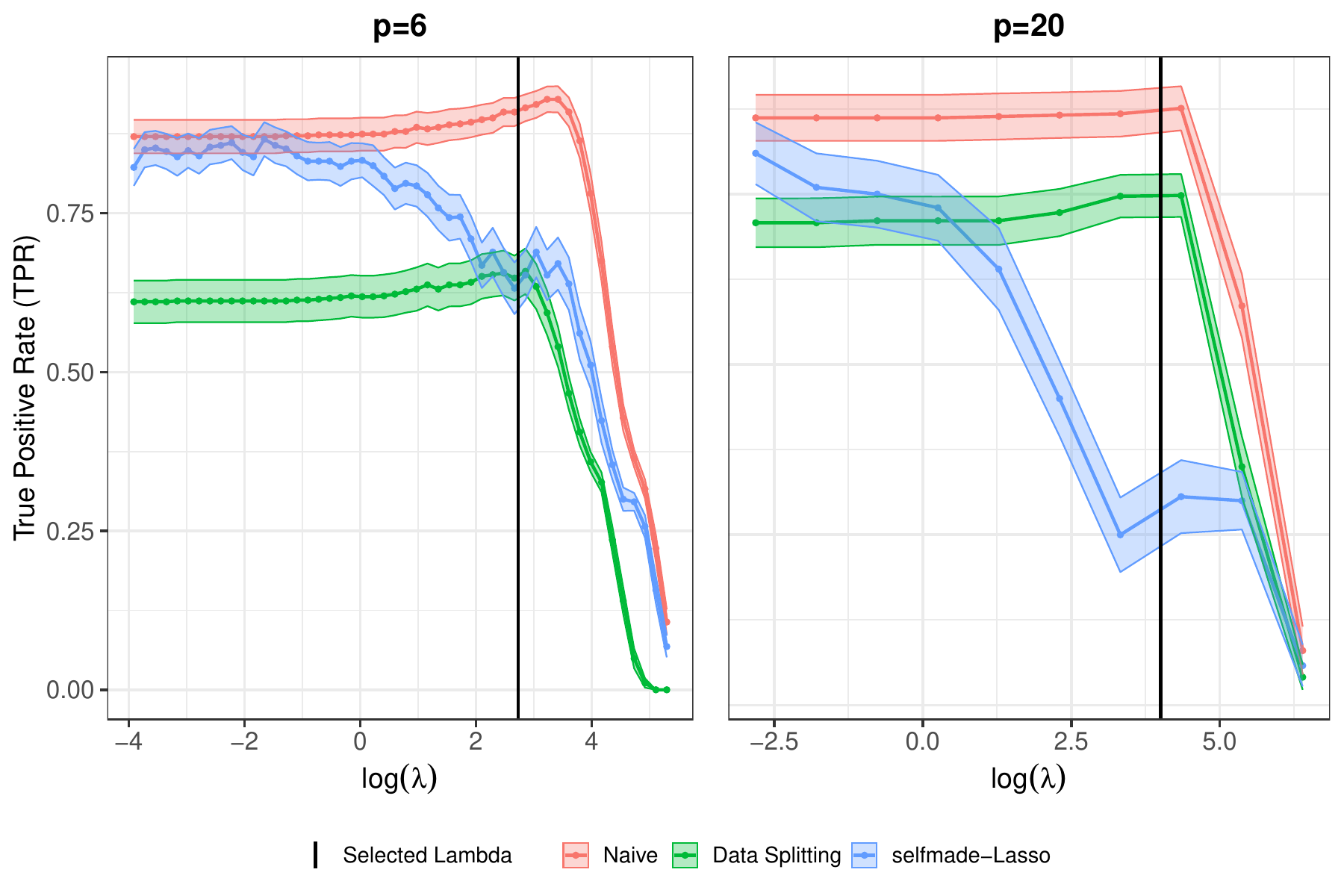}

    \caption{\textit{True positive rate (TPR) as a function of the lasso penalty parameter $\lambda$ for two simulation settings: $p=6$ with 100 observations and $p=20$ with 200 observations. The vertical line represents the average selected $\lambda$ through BIC. The shaded area is a 95\% confidence interval for the average value of TPR out of $50$ simulations.}}
    \label{fig:simulation_changinglambda}
\end{figure}

Figure \ref{fig:simulation_changinglambda} presents the true positive rate (TPR) as a function of the lasso penalty parameter $\lambda$, comparing different inference methods in two scenarios: $ p = 6 $ with 100 observations and $ p = 20 $ with 200 observations. The number of observations is set to ensure comparable initial power levels across both settings. Naïve inference and data splitting exhibit similar trends, with a slight increase in power as $\lambda$ removes noise variables, followed by a sharp decline once signal variables start being excluded. In contrast, selfmade shows a distinct dip in power around the selected $\lambda$, an effect observed for $ p = 6 $ and even more pronounced for $ p = 20 $.
This pattern can once again be explained by looking at the stability of the selection. As mentioned earlier, in the sampling done in Algorithm \ref{alg:rugamer}, only the samples that return the same selection of predictors as the original response are used for inference. When this selection is more stable, such as for low $\lambda$ (where most predictors are retained) or high $\lambda$ (where almost all predictors are excluded), the inference benefits from higher power. However, at intermediate values of $\lambda$, which is typical when model selection is optimized through BIC or cross-validation, the predictor set becomes more unstable. At these values, the inference suffers as fewer samples contribute to the final estimate, resulting in lower power and very large confidence intervals. In light of this, the tuning of the penalization hyperparameter for selfmade (and potentially for other conditional methods) needs to be treated with particular attention.

\section{Real data example}\label{section_real_data_example}

We tested the methods presented on an example dataset, extracted from the Framingham Heart Study. The Framingham Heart Study is a long-term cardiovascular cohort study that has provided data on the impact of lifestyle, genetics, and environmental factors on cardiovascular health \citep{Mahmood2014}. The dataset considered here has first been utilized by \cite{Zhang2001}, and contains: the cholesterol levels of 200 random patients measured at three to six different timepoints over the course of 10 years, the sex of the individuals and their age at the start of the study, and the year of each timepoint. The total number of observations is $1044$.

In order to check the selection results, we generate additional random noise variables, and check what variables are included in the models. The noise variables are distributed normally, with variance computed to obtain a SNR of 0.5. In particular we look at results of FWER and FDR that are computed considering all variables in the initial dataset (sex, age and year) as signal, and the additional random variables as noise. The model is specified as

$$
Y_{i j}= \beta_1 \text{sex}_i+\beta_2 \text{age}_i + \beta_3 t_{i j} + \alpha_{ 1} x_1 + \dots + \alpha_p x_p + b_{0 i}
$$

where $Y_{i j}$ is the cholesterol level (divided by 100 and centered at 0) at the jth time point for the ith subject, sex is a binary variable (0=female, 1=male), age is the subject’s baseline age in years, $t_{i j} = (\tau_{i j} - 5)/10$ where $\tau$ is the time measured in years from the start of the study. $x_1, \dots, x_p$ are the $p$ random noise variables added to the model. $b_{0 i}$ is the random intercept for subject $i$.

The results are obtained by generating the random variables $500$ times.
As in the simulation studies we first consider a simulation with a lower number of variables ($3$ from the dataset and $5$ random noise), in which we look to control the FWER at $5\%$.
The results are presented in Table \ref{table:framingham_small}, which includes the resulting FWER, as well as the average coefficient for sex, age and time, and the percentage of models that include them among the $500$ iterations.  As we can see, the use of naïve inference leads to the violation of constraints on FWER, which is instead controlled by all other methods. As a variable, sex is never included by any method: this is consistent with results from previous analyses of the same dataset which deemed this variable non significant \citep{Galarza2017}. Age and time are almost always included in the models with very similar coefficients: UVILassoLMM is the only method which deviates slightly from these values, as its coefficients are based on the lasso estimator. Data splitting also results in varying coefficient estimates due to the randomness introduced by the data splitting procedure.

The inclusion rates of the age variable vary across methods. The postcAIC consistently selects exactly age and time, whereas selfmade and data splitting exhibit the lowest selection percentages for age.
These differences highlight how different post-selection inference methods can lead to varying analytical outcomes, emphasizing the impact of the selection procedure on inference results.

\begin{table}[]
\resizebox{\textwidth}{!}{
\begin{tabular}{@{}lccccl@{}}
\toprule
\textbf{Method} & \textbf{FWER} & \textbf{Age Avg Coef (\%)} & \textbf{Sex Avg Coef (\%)} & \textbf{Time Avg Coef (\%)} & \multicolumn{1}{c}{\textbf{CI (SD)}} \\ \midrule
Naïve           & 0.104          & 0.118 (91.6)                & / (0)                      & 0.282 (100)                 & 0.058 (0.012)    \\
Data Splitting  & 0.04         & 0.124 (63.2)                & / (0)                      & 0.283 (100)                 & 0.081 (0.018)    \\
postcAIC       & 0             & 0.118 (100)                & / (0)                      & 0.283 (100)                 & 0.106 (0.001)        \\
selfmade-Lasso   & 0.012         & 0.118 (87.2)               & / (0)                      & 0.282 (95.6)                  & 0.078 (0.081)    \\
selfmade-step      & 0.012         & 0.118 (100)               & / (0)                      & 0.283 (99.8)                & 0.062 (0.03)     \\
UVILassoLMM      & 0             & 0.105 (55.6)               & / (0)                      & 0.274 (100)                 & 0.071 (0.012)     \\ \bottomrule
\end{tabular}
}
\caption{\textit{Results for the Framingham dataset with $5$ additional random variables. The table reports the realized FWER and the average estimated coefficient for each variable together with the percentage of models that deemed the coefficient significant out of the $500$ iterations. We also report the average length of confidence intervals and the standard deviation of this value.}}
\label{table:framingham_small}
\end{table}

We also tested the methods that are computationally viable on a larger dataset with $200$ random noise variables. Here, we correct the generated p-values using the Benjamini-Hochberg procedure to control the false discovery rate (FDR) at $5\%$ level. Results are reported in Table \ref{table:framingham_large}. In this setting as well, sex is never selected, while age and time are selected in the majority of cases. Unlike what was observed in the simulation studies, selfmade-Lasso successfully selects signal variables even in the presence of a large number of predictors. Confidence interval lengths vary considerably across methods and, as expected, are generally larger than in the low-dimensional setting. Among the methods considered, selfmade-Lasso returns the widest average confidence intervals.\\

We note that, in order to compute the error metrics (FWER, FDR) we had to make an assumption on which variables were to consider signal and noise. We computed the metrics by considering all added random variables as noise and age, sex and time as signal. In light of the obtained results, and looking at previous analyses, we could have also regarded the sex variable as non relevant. This change in interpretation would not influence the FWER and FDR results in our case, since sex was never included in any model.

\begin{table}[]
\resizebox{\textwidth}{!}{
\begin{tabular}{@{}lccccl@{}}
\toprule
\textbf{Method} & \textbf{FDR} & \textbf{Age Avg Coef (\%)} & \textbf{Sex Avg Coef (\%)} & \textbf{Time Avg Coef (\%)} & \multicolumn{1}{c}{\textbf{CI (SD)}} \\ \midrule
Naïve           & 0.137        & 0.118 (100)                     &   / (0)             & 0.282 (100)                 & \multicolumn{1}{c}{0.078 (0.003)}    \\
Data Splitting  & 0.027        & 0.131 (71.6)                      &    / (0)          & 0.282 (100)                 & 0.115 (0.008)                        \\
selfmade-Lasso   & 0.022        & 0.118 (95.4)                      &    / (0)          & 0.282 (94)                  & 0.292 (0.244)                        \\ \bottomrule
\end{tabular}
}
\caption{\textit{Results for the Framingham dataset as in Table \ref{table:framingham_small}, but with $200$ additional random variables instead.}}
\label{table:framingham_large}
\end{table}

\section{Discussion}\label{section_discussion}

This paper provides a comparative evaluation of various existing methods for selective inference in the context of linear mixed models (LMMs). The methods included consist of sample splitting, an adaptation of the method of confidence regions of \cite{Kramlinger2023}, the conditional inference method of \cite{Rgamer2022} and the cAIC approach of \cite{Claesksen2021}.
After describing them in detail,  we identified key strengths and limitations of these methods through simulations and the analysis of an example real-world dataset taken from the Framingham Heart Study. The code used to perform all the simulations and the real data analysis is available at \url{https://github.com/matteodales/SelectiveInferenceLMMs}.\\

Our comparison with the results of naïve inference, which ignores the effects of selection, reaffirms that this procedure often leads to invalid inferential results, notably inflated type I error rates and reduced coverage probabilities, especially for noise variables. This observation underscores the necessity for methods of inference that account for selection in LMMs even in a low-dimensional setting.
For low-dimensional problems where computational resources are less restrictive, postcAIC provides the most robust results. We note that its implementation is restricted to a nested error regression models (NERM), and does not support more complex random effect structures.
In the presence of additional random effects, we suggest the use of selfmade-step.

For larger models, the only computationally feasible approaches are selfmade, particularly when paired with a selection method like lasso, and sample splitting. Both methods, however, come with drawbacks. Sample splitting inherently sacrifices power, as only part of the data is used for inference, which is especially problematic when the number of observations is limited. The selfmade, on the other hand, suffers from instability in selection as $p$ increases. Because inference is conditioned only on samples that fall within the selection event, unstable selection (where small changes in $\mathbf{y}$ lead to different selected models) results in a substantial proportion of samples being discarded. As a consequence, a much larger number of Monte Carlo samples is required to retain sufficient power, significantly increasing the computational burden. This problem is particularly pronounced in high-dimensional settings, as many selection methods, such as lasso, tend to become more unstable with increasing $p$. The computational complexity of selfmade thus grows not only due to the increased time needed to compute each selection, but also because of the need for an increasingly large number of samples to obtain reliable results.\\

The lack of methodology for LMMs with more than a few fixed effects does not only concern the task of post-selection inference, but even just variable selection with error control. In the context of generalized linear models (GLMs), a range of methods such as knockoffs (\cite{Barber2015}, \cite{Cands2018}) or gaussian mirrors (\cite{Xing2021}) have been developed to provide variable selection with FDR control. However, these approaches typically assume independent observations, making them unsuitable for correlated data structures, such as those modeled by LMMs. Expanding these methods to accommodate the features of LMM data, including clustering and longitudinal dependencies, represents an important area of future research. 

Given the widespread use of LMMs in many areas of research, the issue of selective inference has significant implications for the reliability and reproducibility of scientific findings. Our comparative analysis explores the performance of existing methods in this field and highlights the need for further research that allows their application in a broader group of model settings.

\bibliographystyle{agsm}
\bibliography{sample.bib}

@article{Barber2015,
  title = {Controlling the false discovery rate via knockoffs},
  volume = {43},
  ISSN = {0090-5364},
  DOI = {10.1214/15-aos1337},
  number = {5},
  journal = {The Annals of Statistics},
  publisher = {Institute of Mathematical Statistics},
  author = {Barber,  Rina Foygel and Candès,  Emmanuel J.},
  year = {2015},
  month = oct 
}

@article{Benjamini2005,
  title = {False Discovery Rate–Adjusted Multiple Confidence Intervals for Selected Parameters},
  volume = {100},
  ISSN = {1537-274X},
  DOI = {10.1198/016214504000001907},
  number = {469},
  journal = {Journal of the American Statistical Association},
  publisher = {Informa UK Limited},
  author = {Benjamini,  Yoav and Yekutieli,  Daniel},
  year = {2005},
  month = mar,
  pages = {71–81}
}

@article{Benjamini_2009,
 author = {Benjamini, Yoav and others},
 doi = {10.1098/rsta.2009.0127},
 issn = {1471-2962},
 journal = {Philosophical Transactions of the Royal Society A: Mathematical, Physical and Engineering Sciences},
 month = {November},
 number = {1906},
 pages = {4255–4271},
 publisher = {The Royal Society},
 title = {Selective inference in complex research},
 volume = {367},
 year = {2009}
}

@article{Berk2013,
  title = {Valid post-selection inference},
  volume = {41},
  ISSN = {0090-5364},
  DOI = {10.1214/12-aos1077},
  number = {2},
  journal = {The Annals of Statistics},
  publisher = {Institute of Mathematical Statistics},
  author = {Berk,  Richard and others},
  year = {2013},
  month = apr 
}

@article{Bondell2010,
  title={Joint variable selection for fixed and random effects in linear mixed-effects models},
  author={Bondell, Howard D and others},
  journal={Biometrics},
  volume={66},
  number={4},
  pages={1069--1077},
  year={2010},
  publisher={Wiley Online Library}
}

@article{Cands2018,
  title = {Panning for Gold: ‘{M}odel-{X}’ Knockoffs for High Dimensional Controlled Variable Selection},
  volume = {80},
  ISSN = {1467-9868},
  DOI = {10.1111/rssb.12265},
  number = {3},
  journal = {Journal of the Royal Statistical Society Series B: Statistical Methodology},
  publisher = {Oxford University Press (OUP)},
  author = {Candès,  Emmanuel and Fan,  Yingying and Janson,  Lucas and Lv,  Jinchi},
  year = {2018},
  month = jan,
  pages = {551–577}
}

@article{Charkhi2018,
  title={Asymptotic post‐selection inference for the Akaike information criterion},
  author={Ali Charkhi and Gerda Claeskens},
  journal={Biometrika},
  year={2018},
  volume={105},
  pages={645–664}
}

@article{Claesksen2021,
      title={Post-selection inference for linear mixed model parameters using the conditional Akaike information criterion}, 
      author={Gerda Claeskens and Katarzyna Reluga and Stefan Sperlich},
      year={2021},
journal={arXiv preprint, arXiv:2109.10975}
}

@article{COX1975,
  title = {A note on data-splitting for the evaluation of significance levels},
  volume = {62},
  ISSN = {1464-3510},
  DOI = {10.1093/biomet/62.2.441},
  number = {2},
  journal = {Biometrika},
  publisher = {Oxford University Press (OUP)},
  author = {Cox,  D. R.},
  year = {1975},
  month = aug,
  pages = {441–444}
}

@article{Ewald2018,
  title = {Uniformly valid confidence sets based on the Lasso},
  volume = {12},
  ISSN = {1935-7524},
  DOI = {10.1214/18-ejs1425},
  number = {1},
  journal = {Electronic Journal of Statistics},
  publisher = {Institute of Mathematical Statistics},
  author = {Ewald,  Karl and Schneider,  Ulrike},
  year = {2018},
  month = jan 
}

@article{Fithian2014,
archivePrefix = {arXiv},
arxivId = {arXiv:1410.2597v4},
author = {Fithian, William and Sun, Dennis L. and Taylor, Jonathan},
doi = {10.1198/016214504000001097},
eprint = {arXiv:1410.2597v4},
issn = {01621459},
journal = {arXiv preprint, arXiv:1410.2597},
title = {{Optimal inference after model selection}},
year = {2017}
}

@article{Galarza2017,
  title = {Quantile regression in linear mixed models: a stochastic approximation {EM} approach},
  volume = {10},
  ISSN = {1938-7997},
  DOI = {10.4310/sii.2017.v10.n3.a10},
  number = {3},
  journal = {Statistics and Its Interface},
  publisher = {International Press of Boston},
  author = {Galarza,  Christian E. and others},
  year = {2017},
  pages = {471–482}
}

@article{Ghosh2017,
  title = {Non-concave penalization in linear mixed-effect models and regularized selection of fixed effects},
  volume = {102},
  ISSN = {1863-818X},
  DOI = {10.1007/s10182-017-0298-z},
  number = {2},
  journal = {AStA Advances in Statistical Analysis},
  publisher = {Springer Science and Business Media LLC},
  author = {Ghosh,  Abhik and Thoresen,  Magne},
  year = {2017},
  month = may,
  pages = {179–210}
}

@article{Groll2012,
  title = {Variable selection for generalized linear mixed models by {L}1-penalized estimation},
  volume = {24},
  ISSN = {1573-1375},
  DOI = {10.1007/s11222-012-9359-z},
  number = {2},
  journal = {Statistics and Computing},
  publisher = {Springer Science and Business Media LLC},
  author = {Groll,  Andreas and Tutz,  Gerhard},
  year = {2012},
  month = oct,
  pages = {137–154}
}

@article{Hjort2003,
  title = {Frequentist Model Average Estimators},
  volume = {98},
  ISSN = {1537-274X},
  DOI = {10.1198/016214503000000828},
  number = {464},
  journal = {Journal of the American Statistical Association},
  publisher = {Informa UK Limited},
  author = {Hjort,  Nils Lid and Claeskens,  Gerda},
  year = {2003},
  month = dec,
  pages = {879–899}
}

@article{Ibrahim2010,
  title = {Fixed and Random Effects Selection in Mixed Effects Models},
  volume = {67},
  ISSN = {0006-341X},
  DOI = {10.1111/j.1541-0420.2010.01463.x},
  number = {2},
  journal = {Biometrics},
  publisher = {Oxford University Press (OUP)},
  author = {Ibrahim,  Joseph G. and others},
  year = {2010},
  month = jul,
  pages = {495–503}
}

@article{Jiang2021,
  title = {A generalized linear mixed model association tool for biobank-scale data},
  volume = {53},
  ISSN = {1546-1718},
  DOI = {10.1038/s41588-021-00954-4},
  number = {11},
  journal = {Nature Genetics},
  publisher = {Springer Science and Business Media LLC},
  author = {Jiang,  Longda and others},
  year = {2021},
  month = nov,
  pages = {1616–1621}
}

@article{jones2011bayesian,
  title={Bayesian information criterion for longitudinal and clustered data},
  author={Jones, Richard H},
  journal={Statistics in medicine},
  volume={30},
  number={25},
  pages={3050--3056},
  year={2011},
  publisher={Wiley Online Library}
}

@article{Kramlinger2023,
  title = {Uniformly valid inference based on the Lasso in linear mixed models},
  volume = {198},
  ISSN = {0047-259X},
  DOI = {10.1016/j.jmva.2023.105230},
  journal = {Journal of Multivariate Analysis},
  publisher = {Elsevier BV},
  author = {Kramlinger,  Peter and others},
  year = {2023},
  month = nov,
  pages = {105230}
}

@article{Kubokawa2011,
  title = {Conditional and unconditional methods for selecting variables in linear mixed models},
  volume = {102},
  ISSN = {0047-259X},
  DOI = {10.1016/j.jmva.2010.11.007},
  number = {3},
  journal = {Journal of Multivariate Analysis},
  publisher = {Elsevier BV},
  author = {Kubokawa,  Tatsuya},
  year = {2011},
  month = mar,
  pages = {641–660}
}

@article{Laird_1982,
 author = {Laird, Nan M. and Ware, James H.},
 doi = {10.2307/2529876},
 issn = {0006-341X},
 journal = {Biometrics},
 month = {December},
 number = {4},
 pages = {963},
 publisher = {JSTOR},
 title = {Random-Effects Models for Longitudinal Data},
 volume = {38},
 year = {1982}
}

@article{Lee2016,
  title = {Exact post-selection inference,  with application to the lasso},
  volume = {44},
  ISSN = {0090-5364},
  DOI = {10.1214/15-aos1371},
  number = {3},
  journal = {The Annals of Statistics},
  publisher = {Institute of Mathematical Statistics},
  author = {Lee,  Jason D. and others},
  year = {2016},
  month = jun 
}

@article{Leeb2003,
  title = {THE FINITE-SAMPLE DISTRIBUTION OF POST-MODEL-SELECTION ESTIMATORS AND UNIFORM VERSUS NONUNIFORM APPROXIMATIONS},
  volume = {19},
  ISSN = {1469-4360},
  DOI = {10.1017/s0266466603191050},
  number = {01},
  journal = {Econometric Theory},
  publisher = {Cambridge University Press (CUP)},
  author = {Leeb,  Hannes and P\"{o}tscher,  Benedikt M.},
  year = {2003},
  month = jan 
}

@article{Leeb2006,
  title = {Can one estimate the conditional distribution of post-model-selection estimators?},
  volume = {34},
  ISSN = {0090-5364},
  DOI = {10.1214/009053606000000821},
  number = {5},
  journal = {The Annals of Statistics},
  publisher = {Institute of Mathematical Statistics},
  author = {Leeb,  Hannes and P\"{o}tscher,  Benedikt M.},
  year = {2006},
  month = oct 
}

@article{Leeb2007,
  title = {CAN ONE ESTIMATE THE UNCONDITIONAL DISTRIBUTION OF POST-MODEL-SELECTION ESTIMATORS?},
  volume = {24},
  ISSN = {1469-4360},
  DOI = {10.1017/s0266466608080158},
  number = {02},
  journal = {Econometric Theory},
  publisher = {Cambridge University Press (CUP)},
  author = {Leeb,  Hannes and P\"{o}tscher,  Benedikt M.},
  year = {2007},
  month = nov 
}

@article{Lindstrom1988,
  title = {Newton-Raphson and EM Algorithms for Linear Mixed-Effects Models for Repeated-Measures Data},
  volume = {83},
  ISSN = {0162-1459},
  DOI = {10.2307/2290128},
  number = {404},
  journal = {Journal of the American Statistical Association},
  publisher = {JSTOR},
  author = {Lindstrom,  Mary J. and Bates,  Douglas M.},
  year = {1988},
  month = dec,
  pages = {1014}
}

@article{Loftus2015,
      title={Selective inference after cross-validation}, 
      author={Loftus,  Joshua R.},
      year={2015},
journal={arXiv preprint, arXiv:1511.08866}
}

@article{Mahmood2014,
  title = {The {F}ramingham {H}eart {S}tudy and the epidemiology of cardiovascular disease: a historical perspective},
  volume = {383},
  ISSN = {0140-6736},
  DOI = {10.1016/s0140-6736(13)61752-3},
  number = {9921},
  journal = {The Lancet},
  publisher = {Elsevier BV},
  author = {Mahmood,  Syed S and others},
  year = {2014},
  month = mar,
  pages = {999–1008}
}

@article{Meinshausen_2009,
 author = {Meinshausen, Nicolai and others},
 doi = {10.1198/jasa.2009.tm08647},
 issn = {1537-274X},
 journal = {Journal of the American Statistical Association},
 month = {December},
 number = {488},
 pages = {1671–1681},
 publisher = {Informa UK Limited},
 title = {p-Values for High-Dimensional Regression},
 volume = {104},
 year = {2009}
}

@article{Ptscher1991,
  title = {Effects of Model Selection on Inference},
  volume = {7},
  ISSN = {1469-4360},
  DOI = {10.1017/s0266466600004382},
  number = {2},
  journal = {Econometric Theory},
  publisher = {Cambridge University Press (CUP)},
  author = {P\"{o}tscher,  B.M.},
  year = {1991},
  month = jun,
  pages = {163–185}
}

@article{Rgamer2022,
  title = {Selective inference for additive and linear mixed models},
  volume = {167},
  ISSN = {0167-9473},
  DOI = {10.1016/j.csda.2021.107350},
  journal = {Computational Statistics \& Data Analysis},
  publisher = {Elsevier BV},
  author = {R\"{u}gamer,  David and others},
  year = {2022},
  month = mar,
  pages = {107350}
}

@article{Schelldorfer2014,
  title = {{GLMML}asso: An Algorithm for High-Dimensional Generalized Linear Mixed Models Using {L}1-Penalization},
  volume = {23},
  ISSN = {1537-2715},
  DOI = {10.1080/10618600.2013.773239},
  number = {2},
  journal = {Journal of Computational and Graphical Statistics},
  publisher = {Informa UK Limited},
  author = {Schelldorfer,  J\"{u}rg and others},
  year = {2014},
  month = apr,
  pages = {460–477}
}

@article{Sugasawa2020,
  title = {Small area estimation with mixed models: a review},
  volume = {3},
  ISSN = {2520-8764},
  DOI = {10.1007/s42081-020-00076-x},
  number = {2},
  journal = {Japanese Journal of Statistics and Data Science},
  publisher = {Springer Science and Business Media LLC},
  author = {Sugasawa,  Shonosuke and Kubokawa,  Tatsuya},
  year = {2020},
  month = apr,
  pages = {693–720}
}

@article{Tian2018,
  title = {Selective inference with a randomized response},
  volume = {46},
  ISSN = {0090-5364},
  DOI = {10.1214/17-aos1564},
  number = {2},
  journal = {The Annals of Statistics},
  publisher = {Institute of Mathematical Statistics},
  author = {Tian,  Xiaoying and Taylor,  Jonathan},
  year = {2018},
  month = apr 
}

@article{Tibshirani2016,
  title = {Exact Post-Selection Inference for Sequential Regression Procedures},
  volume = {111},
  ISSN = {1537-274X},
  DOI = {10.1080/01621459.2015.1108848},
  number = {514},
  journal = {Journal of the American Statistical Association},
  publisher = {Informa UK Limited},
  author = {Tibshirani,  Ryan J. and others},
  year = {2016},
  month = apr,
  pages = {600–620}
}

@article{Vaida2005,
  title = {Conditional {A}kaike information for mixed-effects models},
  volume = {92},
  ISSN = {0006-3444},
  DOI = {10.1093/biomet/92.2.351},
  number = {2},
  journal = {Biometrika},
  publisher = {Oxford University Press (OUP)},
  author = {Vaida,  Florin and Blanchard,  Suzette},
  year = {2005},
  month = jun,
  pages = {351–370}
}

@article{vandeGeer2014,
  title = {On asymptotically optimal confidence regions and tests for high-dimensional models},
  volume = {42},
  ISSN = {0090-5364},
  DOI = {10.1214/14-aos1221},
  number = {3},
  journal = {The Annals of Statistics},
  publisher = {Institute of Mathematical Statistics},
  author = {van de Geer,  Sara and others},
  year = {2014},
  month = jun 
}

@article{Wald1943,
  title = {Tests of statistical hypotheses concerning several parameters when the number of observations is large},
  volume = {54},
  ISSN = {1088-6850},
  DOI = {10.1090/s0002-9947-1943-0012401-3},
  number = {3},
  journal = {Transactions of the American Mathematical Society},
  publisher = {American Mathematical Society (AMS)},
  author = {Wald,  Abraham},
  year = {1943},
  pages = {426–482}
}

@article{Xing2021,
  title = {Controlling False Discovery Rate Using Gaussian Mirrors},
  volume = {118},
  ISSN = {1537-274X},
  DOI = {10.1080/01621459.2021.1923510},
  number = {541},
  journal = {Journal of the American Statistical Association},
  publisher = {Informa UK Limited},
  author = {Xing,  Xin and Zhao,  Zhigen and Liu,  Jun S.},
  year = {2021},
  month = jun,
  pages = {222–241}
}

@ARTICLE{Xu2012,
  author={Xu, Huan and Caramanis, Constantine and Mannor, Shie},
  journal={IEEE Transactions on Pattern Analysis and Machine Intelligence}, 
  title={Sparse Algorithms Are Not Stable: A No-Free-Lunch Theorem}, 
  year={2012},
  volume={34},
  number={1},
  pages={187-193},
  doi={10.1109/TPAMI.2011.177}}

@inproceedings{Yang2016,
 author = {Yang, Fan and others},
 booktitle = {\textit{Advances in Neural Information Processing Systems}},
 pages = {2469-2477},
 title = {Selective inference for group-sparse linear models},
 volume = {29},
 year = {2016}
}

@article{Zhang2001,
  title = {Linear Mixed Models with Flexible Distributions of Random Effects for Longitudinal Data},
  volume = {57},
  ISSN = {0006-341X},
  DOI = {10.1111/j.0006-341x.2001.00795.x},
  number = {3},
  journal = {Biometrics},
  publisher = {Oxford University Press (OUP)},
  author = {Zhang,  Daowen and Davidian,  Marie},
  year = {2001},
  month = sep,
  pages = {795–802}
}

@article{Zhang2022,
  title = {Post-model-selection inference in linear regression models: An integrated review},
  volume = {16},
  ISSN = {1935-7516},
  DOI = {10.1214/22-ss135},
  number = {none},
  journal = {Statistics Surveys},
  publisher = {Institute of Mathematical Statistics},
  author = {Zhang,  Dongliang and others},
  year = {2022},
  month = jan 
}

@article{Holm1979,
  title={A Simple Sequentially Rejective Multiple Test Procedure},
  author={Sture Holm},
  journal={Scandinavian Journal of Statistics},
  year={1979},
  volume={6},
  pages={65-70}
}

@article{Benjamini1995,
  title = {Controlling the False Discovery Rate: A Practical and Powerful Approach to Multiple Testing},
  volume = {57},
  ISSN = {1467-9868},
  DOI = {10.1111/j.2517-6161.1995.tb02031.x},
  number = {1},
  journal = {Journal of the Royal Statistical Society Series B: Statistical Methodology},
  publisher = {Oxford University Press (OUP)},
  author = {Benjamini,  Yoav and Hochberg,  Yosef},
  year = {1995},
  month = jan,
  pages = {289–300}
}

\end{document}